\documentclass[aps,prl,twocolumn,superscriptaddress,preprintnumbers,amsmath,amssymb,showkeys,floatfix,nofootinbib,reprint]{revtex4-1}

\usepackage{amsfonts}
\usepackage{amssymb}
\usepackage{graphics}
\usepackage{graphicx}
\usepackage{epstopdf}
\usepackage{dcolumn}
\usepackage{bm}
\usepackage{longtable}
\usepackage{epsfig}
\usepackage{times}
\usepackage{url}
\usepackage{color}
\usepackage{amsmath}
\usepackage{comment}
\usepackage{mathtools}

\begin{document}

\title{Nuclear excitation by radiative electron-ion recombination}

\author{Jingyan \surname{Zhao}}
\affiliation{School of Physics, Nankai University, Tianjin 300071, China}

\author{Yuanbin \surname{Wu}}
\email{yuanbin@nankai.edu.cn}
\affiliation{School of Physics, Nankai University, Tianjin 300071, China}

\date{\today}

\begin{abstract}

A nuclear excitation mechanism, nuclear excitation by radiative electron-ion recombination (NERER), is put forward theoretically here. NERER is a third-order process that proceeds via a virtual electronic state: an electron recombines into an atomic vacancy of an ion with the simultaneous emission of a real photon and excitation of the nucleus. The photon emission compensates the energy mismatch between the free-bound electronic transition and the nuclear transition energies, thus there is no resonant condition imposed to the incident electron. We develop here the theoretical framework for NERER, and investigate the case of the $8.4$ eV isomeric excitation of $^{229}$Th for the production of the nuclear clock isomer $^{229m}$Th. Our results show that, with the coupling to the inner atomic shells for highly-charged ions, the NERER cross section can exceed the one of the known lower-order process of nuclear excitation by inelastic electron scattering by more than one order of magnitude. Our findings offer a new pathway for nuclear excitation and efficient isomer production, and support further investigations for high-order effects in the interplay between the atomic and nuclear systems.

\end{abstract}

\maketitle


Long-lived excited states of atomic nuclei, known as nuclear isomers, have attracted considerable attention, due to their importance in studies in nuclear theory, as well as potential applications such as nuclear energy storage, nuclear $\gamma$-ray lasers, nuclear clocks, and nuclear medicine \cite{Walker1999N,Walker2020PS,Dracoulis2016RPP,Palffy2007PRL,Carroll2007NIMPRB,Tkalya2011PRL,vonderWense2020EPJA,Peik2003EL,Burke2019N,Beeks2021NRP,Peik2021QST}. One of the key issues is the means of the efficient and controllable production and depletion of isomers. Driving nuclear transitions via the electron-induced nuclear transition processes is believed to be one of the promising methods. Numerous electron-induced nuclear excitation mechanisms have been proposed, including nuclear excitation by inelastic electron scattering (NEIES) \cite{Thie1952PR,Schiff1954PR,Liu2022PRC,Harston1999PRC,Tkalya2020PRL,Zhang2022PRC,Zhang2023FP,Xu2024PRC,Wang2026PRC,Feng2022PRL,Feng2024PNAS}, nuclear excitation by electron capture (NEEC) \cite{Goldanskii1976PLB,Doolen1978PRL,Harston1999PRC,Gosselin2004PRC,Palffy2006PRA,Palffy2007PRA,Palffy2007PRL,Gunst2014PRL,Wu2018PRL,Cerjan2018JPGNPP,Chiara2018N,Wu2022PRL,Gargiulo2022PRL,MA2022SB,Spohr2023EPJA,Qi2023PRL}, nuclear excitation by electron transition (NEET) \cite{Morita1973PTP,Tkalya1992NPA,Kishimoto2000PRL,Kishimoto2005NPA,Morel2010PRC,Chodash2016PRC,Karpeshin1996PLB,Karpeshin2017PRC,Koziol2025PRC}, and processes with the absorption of additional photons such as electronic bridge (EB) \cite{Porsev2010PRL,Berengut2018PRL,Bilous2020PRL,Nickerson2020PRL,Wang2024PRL,Borisyuk2019PRC,Dzyublik2020PRC,Dzuba2025PRA} and other laser-assisted electron-nucleus processes \cite{Xu2023PRA,Karpeshin2024PRC}. One of the intriguing points here is the coupling between the atomic and nuclear degrees of freedom.

An exciting case is the isomer $^{229m}$Th \cite{vonderWense2020EPJA,Peik2003EL,Beeks2021NRP,Burke2019N,Peik2021QST,Tiedau2024PRL,Elwell2024PRL,Zhang2024N633,Zhang2024N636,Higgins2025PRL,Elwell2025N,Ooi2026N}, the lowest known metastable nuclear excited state with an excitation energy of approximately $8.4$ eV which makes $^{229}$Th as a crucial candidate for nuclear clocks with an extremely high accuracy \cite{vonderWense2020EPJA,Peik2003EL,Burke2019N,Beeks2021NRP,Campbell2012PRL,Kazakov2012NJP,Peik2021QST,Rellergert2010PRL}. With this unique isomer, $^{229}$Th has also been shown to be able to offer new opportunities for many prospects such as the detection of dark mater and temporal variations of fundamental constants, and studies of effects of the electronic environment on nuclear properties \cite{Beeks2021NRP,Peik2021QST,Flambaum2006PRL,Berengut2009PRL,Fadeev2022PRC,Beeks2025NC,Morgan2025PRL,Flambaum2016PRL,Shabaev2022PRL,Perera2025PRL,Zhou2026PRL}. After sustained efforts to populate $^{229m}$Th and measure its properties via a number of methods \cite{Beck2007PRL,vonderWense2016N,Seiferle2017PRL,Thielking2018N,Seiferle2019N,Yamaguchi2019PRL,Sikorsky2020PRL,Yamaguchi2024N,Kraemer2023N,Masuda2019N}, direct laser excitation to populate $^{229m}$Th has been experimental demonstrated \cite{Tiedau2024PRL,Elwell2024PRL,Zhang2024N633,Zhang2024N636}, and very recently first nuclear clocks have been demonstrated \cite{Col2026arXiv,Huang2026arXiv}. In the context of the coupling between the atomic and nuclear degrees of freedom, the production of $^{229m}$Th via the process of NEIES \cite{Tkalya2020PRL,Zhang2022PRC,Xu2024PRC,Wang2026PRC}, NEEC \cite{Qi2023PRL,Zhao2024PRC,Wang2026PRC,xu2025arXiv}, NEET \cite{Karpeshin1996PLB,Karpeshin2017PRC,Koziol2025PRC}, and laser-assisted electron-nucleus processes \cite{Porsev2010PRL,Bilous2020PRL,Nickerson2020PRL,Dzyublik2020PRC,Wang2024PRL,Borisyuk2019PRC,Dzuba2025PRA,Xu2023PRA,Karpeshin2024PRC} such as EB have been investigated theoretically. Measurements have also shown that the lifetime of $^{229m}$Th is very sensitive to the charge state of the ion \cite{Seiferle2017PRL,Thielking2018N,vonderWense2016N,Yamaguchi2024N,Shigekawa2026NP}, indicating the importance of the electron-induced nuclear de-excitation. Moreover, recent experiments on the photon-/laser-induced quenching of $^{229m}$Th \cite{Hiraki2024NC,Terhune2025PRR,Schaden2025PRR} have also demonstrated the coupling between the atomic and nuclear systems. We note that NEET and EB require and rely on the detailed and precise properties of the excited atomic states, and the preparation of the excited atomic states or the match between the laser/additional photon and the atomic states.

NEIES and NEEC are among the most popular mechanisms of nuclear excitation induced by electron sources. In NEIES, the nuclear excitation is driven by the electron transition from a higher-energy continuum state to a lower-energy continuum state. With the free-free electronic transition, there is no resonant condition imposed to the incident electron for NEIES, and recent experiments have shown that NEIES could be a potential tool for isomer production \cite{Feng2022PRL,Feng2024PNAS}. In contrast, NEEC, in which the nuclear excitation is driven by the capture of an electron into an atomic shell, is a resonant process which requires the match between the free-bound electronic transition and nuclear transition energies. It has been shown that NEEC may play important roles in various aspects \cite{Palffy2007PRL,Gunst2014PRL,Wu2018PRL,Cerjan2018JPGNPP,Qi2023PRL} including isomer depletion and the production of the nuclear clock isomer $^{229m}$Th. Recently, NEEC has attracted a great deal of attention due to the discussions and contradictory results \cite{Wu2019PRL,Rzadkiewicz2021PRL,Rzadkiewicz2023PRC,Guo2021N,Chiara2021N,Guo2022PRL,Ding2026PRL} from both theoretical and experimental aspects on the first reported experimental evidence \cite{Chiara2018N}. The efficient and controllable production and depletion of isomers, and the identification of intriguing phenomena of the interplay between the atomic and nuclear degrees of freedom, are among the current main focuses in nuclear physics and atomic physics. Investigation on possible new nuclear excitation mechanisms with high efficiencies which could surpass the ones of the known mechanisms is highly demanded.

\begin{figure}[htb!]
    \centering
    \includegraphics[width=1\linewidth]{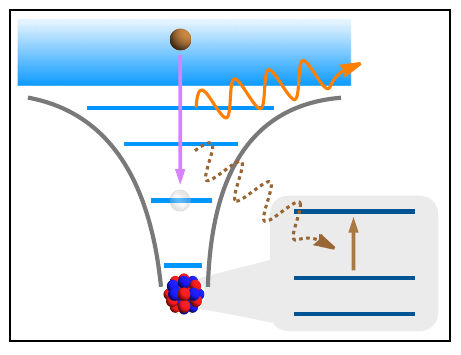}
    \caption{Sketch of nuclear excitation by radiative electron-ion recombination.}
    \label{figsketch}
\end{figure}

We put forward here a nuclear excitation mechanism: nuclear excitation by radiative electron-ion recombination (NERER). In NERER, as shown in Fig.~\ref{figsketch}, the nuclear excitation is driven by an electron recombining into an atomic vacancy of an ion with the simultaneous emission of a real photon. NERER is a third-order process that proceeds via a virtual electronic state. The photon emission compensates the energy mismatch between the free-bound electronic transition and the nuclear transition energies, thus there is no resonant condition imposed to the incident electron. We develop here the theoretical approach for NERER. As an interesting example, we investigate numerically the case of the $8.4$ eV isomeric excitation of $^{229}$Th. Our calculations show that, with the coupling to the inner atomic shells for highly-charged ions, the NERER cross section can exceed the one of the known lower-order process of NEIES by more than one order of magnitude. Our findings offer a new pathway for the efficient production of the nuclear clock isomer $^{229m}$Th, and highlight the importance of high-order effects.

The initial state $\left|\psi_{i}\right>$ and the final state $\left|\psi_{f}\right>$ of the system can be written as $\left|\psi_{i}\right>=\left|I_{i}M_{I_{i}}\right>\otimes\left|\vec{p}_{i}m_{s}\right>\otimes\left|0\right>$ and $\left|\psi_{f}\right> = \left|I_{f}M_{I_{f}}\right>\otimes\left|n_{f}\kappa_{f}m_{f}\right>\otimes\left|\lambda_{f}k_{f}L_{f}M_{f}\right>$. Here, $\left|I_{i,f}M_{I_{i,f}}\right>$ is the initial or final state of the nucleus with the total angular momentum quantum number $I_{i,f}$ and the magnetic quantum number $M_{I_{i,f}}$. $\left|\vec{p}_{i}m_{s}\right>$ is the free-electron state with the asymptotic momentum $\vec{p}_{i}$ and spin projection $m_{s}$. $\left|n_{f}\kappa_{f}m_{f}\right>$ is the final state of the bound electron with the principal quantum number $n_{f}$, the Dirac angular momentum quantum number $\kappa_{f}$ and the magnetic quantum number $m_{f}$. $\left|\lambda_{f}k_{f}L_{f}M_{f}\right>$ is the state of the emitted photon with $\lambda_{f}$ standing for electric $(\lambda_{f}\!=\!E)$ or magnetic $(\lambda_{f}\!=\!M)$ waves, the wave number $k_{f}$, the total angular momentum $L_{f}$ and its projection $M_{f}$. $\left|0\right>$ is the vacuum state of the electromagnetic field. Following similar approaches for NEEC and NEET in Refs.~\cite{Palffy2006PRA,Palffy2007PRA,Arigapudi2012PRA}, and performing the average over the initial states and the summation over the final states, the cross section for NERER for a given recombined electron orbital is given by
\begin{align}
    \sigma_{f}
    =&\frac{1}{2}\frac{1}{2I_{i}+1}\sum_{M_{I_{i}}m_{s}}\sum_{M_{I_{f}}m_{f}}\sum_{\lambda_{f}L_{f}M_{f}}\frac{1}{4\pi}\int\mathrm{d}\Omega_{p_{i}}
    \nonumber\\
    &\times\frac{2\pi}{F_{i}}\lim_{\epsilon\rightarrow 0^{+}}\left|\left<\psi_{f}\right|T\left(E+i\epsilon\right)\left|\psi_{i}\right>\right|^{2}\rho\left(E_{\mathrm{ph}}\right),
\end{align}
where $\rho\left(E_{\mathrm{ph}}\right)$ is the density of the final photonic states, and $T\left(E+i\epsilon\right)$ is the transition operator. $\Omega_{p_{i}}$ and $F_{i}$ denote the solid angle and the flux of the incident electron. By developing the Feshbach projection operator for the process and employing the Lippmann-Schwinger equation to obtain the perturbation series for the transition operator, we obtain
\begin{align} \label{eq:sigmaF}
    \sigma_{f}
    &=\frac{1}{2}\frac{1}{2I_{i}+1}\sum_{M_{I_{i}}m_{s}}\sum_{M_{I_{f}}m_{f}}\sum_{\lambda_{f}L_{f}M_{f}}\frac{1}{4\pi}\int\mathrm{d}\Omega_{p_{i}}
    \nonumber\\
    &\times\frac{2\pi}{F_{i}}\left(\left|F_{1}\right|^{2}+\left|F_{2}\right|^{2}+
    2\mathrm{Re}\left(F_{1}F_{2}^{*}\right)
    \right)\rho\left(E_{\mathrm{ph}}\right),
\end{align}
where
\begin{align} \label{eq:sFone}
    F_{1} &= \!\! \sum_{n_{1}\kappa_{1}m_{1}} \frac{\left<N_{f}I_{f}M_{I_{f}};n_{f}\kappa_{f}m_{f}\right|H_{\mathrm{int}}\left|N_{i}I_{i}M_{I_{i}};n_{1}\kappa_{1}m_{1}\right>}{E_{f}^{n}+E_{f}^{e}-E_{1}^{e}-E_{i}^{n}+i\Gamma_{1}/2}\nonumber\\
    & \times\left<n_{1}\kappa_{1}m_{1};\lambda_{f}k_{f}L_{f}M_{f}\right|H_{er}\left|\vec{p}_{i}m_{s}\right>,
\end{align}
and
\begin{align} \label{eq:sFtwo}
    F_{2} &= \sum_{n_{2}\kappa_{2}m_{2}}\frac{\left<n_{f}\kappa_{f}m_{f};\lambda_{f}k_{f}L_{f}M_{f}\right|H_{er}\left|n_{2}\kappa_{2}m_{2}\right>}{E_{i}^{n}+E_{i}^{e}-E_{f}^{n}-E_{2}^{e}+i\Gamma_{2}/2}
    \nonumber\\
    &\times \left<N_{f}I_{f}M_{I_{f}};n_{2}\kappa_{2}m_{2}\right|H_{\mathrm{int}}\left|N_{i}I_{i}M_{I_{i}};\vec{p}_{i}m_{s}\right>\nonumber\\
    &+
    \int\frac{\mathrm{d}\vec{p}}{\left(2\pi\right)^{3}}\sum_{\mu}\frac{\left<n_{f}\kappa_{f}m_{f};\lambda_{f}k_{f}L_{f}M_{f}\right|H_{er}\left|\vec{p}\mu\right>}{E_{i}^{e}+E_{i}^{n}-\left(E_{f}^{n}+E^{e}_{p}\right)+i\Gamma_{3}/2}\nonumber\\
    & \times\left<N_{f}I_{f}M_{I_{f}};\vec{p}\mu\right|H_{\mathrm{int}}\left|N_{i}I_{i}M_{I_{i}};\vec{p}_{i}m_{s}\right>
    .
\end{align}
Here $H_{\mathrm{int}}=H_{en}+H_{magn}$ with the electric interaction Hamiltonian $H_{en}$ and the magnetic interaction Hamiltonian $H_{magn}$ for the electron-nucleus interaction, and $H_{er}$ is Hamiltonian of the interaction between the electron and the radiation field. $E^{n}$ is the energy of the nucleus, $E^{e}$ is the kinetic energy of the electron, and $\Gamma$ is the width of the state. The details of the Hamiltonian can be found in Refs.~\cite{Palffy2006PRA,Palffy2007PRA,Arigapudi2012PRA}. The process can be decomposed into two paths: (i) the electron transits from the initial state to a virtual state with a photon emission followed by the nuclear excitation by the electron transition from the virtual state to the final electronic state, and (ii) the nucleus is excited by the electron transition from the initial state to a virtual state followed by the electron transition from the virtual state to the final electronic state with a photon emission. To calculate $F_1$ and $F_2$, summation over the discrete electron states $\left|n_1 \kappa_1 m_1\right>$ or $\left|n_2 \kappa_2 m_2\right>$ and integration over the continuum electron spectrum $\left|\vec{p} \mu \right>$ are required. However, as one of our purposes is to study the effect of the coupling of the inner atomic shells, we ignore in the present work the contribution from the continuum electron spectrum $\left|\vec{p} \mu \right>$ for $F_1$, as shown in Eq.~\eqref{eq:sFone}. Such term may play important roles when the nuclear transition energy exceeds the binding energy of the recombined electron orbital.

In order to calculate the matrix elements, we expand the free electron wave function $\left|\vec{p}_{i}m_{s}\right>$ into partial wave series \cite{Zhang2022PRC,Liu2022PRC,Zhang2023FP,Rose1961AJP}, and we obtain
\begin{equation} \label{eq:sigG}
  \sigma_{f} = \frac{1}{4(2I_{i}+1)F_i}\rho\left(E_{\mathrm{ph}}\right)
    \left[
    G_{1}+G_{2}+2\mathrm{Re}(G_{12})
    \right],
\end{equation}
with
\begin{align}
		G_{1} &= \sum_{m_{s}M_{I_{i}}}\sum_{\lambda_{f}L_{f}M_{f}}\sum_{M_{I_{f}}m_{f}} \int\mathrm{d}\Omega_{p_{i}}\left|F_{1}\right|^{2},
\end{align}
\begin{align}
        G_{2} &= \sum_{m_{s}M_{I_{i}}}\sum_{\lambda_{f}L_{f}M_{f}}\sum_{M_{I_{f}}m_{f}} \int\mathrm{d}\Omega_{p_{i}}\left|F_{2}\right|^{2}, \\
        G_{12}
        & = \sum_{m_{s}M_{I_{i}}}\sum_{\lambda_{f}L_{f}M_{f}}\sum_{M_{I_{f}}m_{f}} \int\mathrm{d}\Omega_{p_{i}}F_{1}F_{2}^{*}.
\end{align}
The details of $G_1$, $G_2$, and $G_{12}$ can be found in the end matter. As presented in the end matter, the transition matrix elements include the electronic part, and the nuclear part which can be connected to the nuclear reduced transition probabilities.


As an interesting and important example, we focus in the following on the case of the nuclear transition of $^{229}$Th from the ground state to the first excited state $^{229m}$Th. We adopt the nuclear reduced transition probability $B\left(E2\right)=27$ W.u. (Weisskopf units) from the theoretical calculation in Ref.~\cite{Minkov2017PRL} and $B\left(M1\right)=0.022$ W.u. from the measurement in Th-doped crystals in Ref.~\cite{Tiedau2024PRL}, for the nuclear transition from the isomeric state $^{229m}$Th to the ground state. We note that the discussion on the uncertainties of the nuclear reduced transition probabilities for the radiative decay of $^{229m}$Th can be found in Ref.~\cite{Chen2025PLB}. The excitation energy of $^{229m}$Th is adopted from Ref.~\cite{ensdf2026}, which is approximately $8.36$ eV. In order to calculate the relativistic radial wave functions of electrons, we adopt the code RADIAL \cite{Salvat2019CPC} with the Dirac-Hartree-Fock-Slater method \cite{Liberman1965PR,Liberman1971CPC} and a Fermi charge distribution for the charge density of the nucleus.

\begin{figure}[htb!]
    \centering
    \includegraphics[width=1\linewidth]{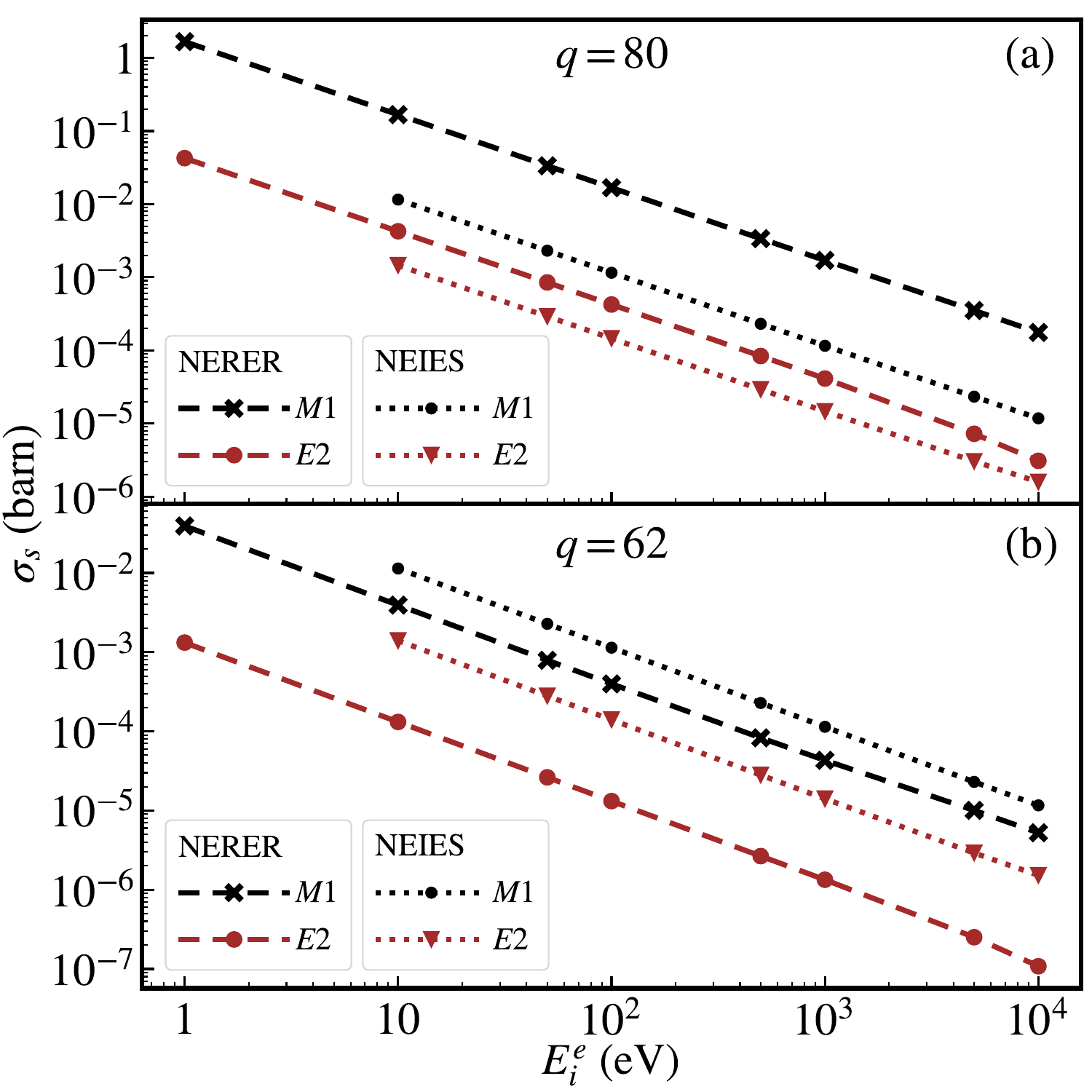}
    \caption{The summation cross section $\sigma_{s}$ of NERER and the NEIES cross section as functions of the incident electron energy $E^e_i$ for the initial charge state of $^{229}$Th of (a) $q=80$ and (b) $q=62$. For $q=80$, $\sigma_{s}$ is defined as the summation of the NERER cross sections over all channels with the final electron states of $M$ and $N$ shells, while $\sigma_{s}$ is defined as the summation over all channels with the final electron states of $N$ and $O$ shells for $q=62$.}
    \label{fig:sigVsNEIES}
\end{figure}

As there is no resonant condition imposed to the incident electron in NERER, for a given energy of the incident electron, the process can occur for the recombined orbitals with the energy exceeds the threshold, i.e., the incident electron energy plus the binding energy of the recombined electron orbital is larger than the nuclear excitation energy. We analyse at first the summation cross section $\sigma_{s}$ of NERER, which is the summation of the NERER cross sections over certain recombined channels of inner atomic shells. Figure~\ref{fig:sigVsNEIES} shows the results for the summation cross section $\sigma_{s}$ of NERER as functions of the incident electron energy. Low energy electrons are considered. Here two initial charge states of $^{229}$Th of $q=80$ and $q=62$ are considered, and the initial state is assumed to be in its electronic ground state of the given charge state of the ion. This means that $K$ and $L$ shells are fully occupied in the initial state of the ion for $q=80$, while $K$, $L$, and $M$ shells are fully occupied in the initial state of the ion for $q=62$. For the case of $q=80$, the summation cross section $\sigma_{s}$ of NERER is defined as $\sigma_s = \sum_{n_f = 3}^{4} \sigma_f$. And for the case of $q=62$, $\sigma_{s}$ is defined as $\sigma_s = \sum_{n_f = 4}^{5} \sigma_f$. The $M1$ nuclear transition and the $E2$ nuclear transition are separately presented in Fig.~\ref{fig:sigVsNEIES}. It is shown in Fig.~\ref{fig:sigVsNEIES} that the summation cross section $\sigma_{s}$ of NERER decreases with increasing the energy of the incident electron. Figure~\ref{fig:sigVsNEIES} also shows a strong dependence of the NERER cross section on the charge state: for each nuclear transition multipolarity, $\sigma_{s}$ for $q=80$ is larger than the one for $q=62$ at each incident electron energy considered here. This implies the effect of the coupling to the inner atomic shell for the nuclear excitation.

For the sake of comparisons, we also calculate the NEIES cross sections following the method in Refs.~\cite{Zhang2022PRC,Li2026PRC}. The results are presented in Fig.~\ref{fig:sigVsNEIES}. We can observe in Fig.~\ref{fig:sigVsNEIES} that for the case of $q=80$, the NERER cross section can exceed the one of the lower-order process of NEIES by more than one order of magnitude. This result shows clear the strong effect of the coupling to the inner atomic shells for highly-charged ions on the nuclear excitation, in which the electron cloud is close to the nucleus. And our result also highlights the importance of high-order effects in the interplay between the atomic and nuclear degrees of freedom. 

We note that NEIES occurs when the energy of the incident electron exceeds the nuclear transition energy. This is different from the case of NERER, in which the threshold condition is that the incident electron energy plus the binding energy of the recombined electron orbital is larger than the nuclear excitation energy. This is also clearly shown in Fig.~\ref{fig:sigVsNEIES} that NERER can occur when the energy of the incident electron is less than the excitation energy of $^{229m}$Th, while NEIES is forbidden. It is also shown in Fig.~\ref{fig:sigVsNEIES} that the cross section of NERER exceeds $1$ barn for low electron energy around $1$ eV. This result shows that NERER could be a potential pathway for the efficient production of the nuclear clock isomer $^{229m}$Th.

\begin{figure}[htb!]
\centering    
\includegraphics[width=1\linewidth]{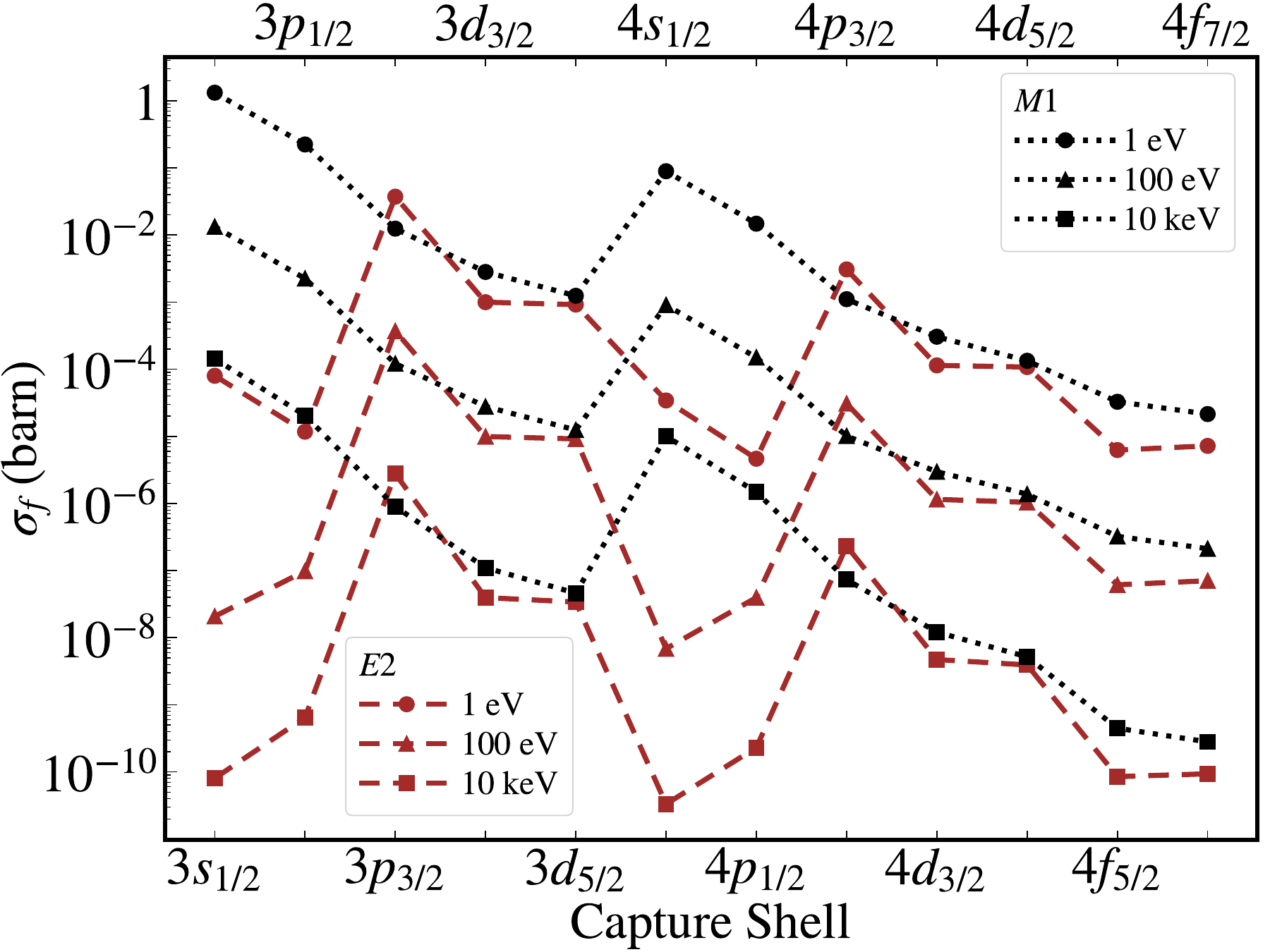}
\caption{The NERER cross section $\sigma_f$ for each recombined electron orbital of $M$ and $N$ shells for the initial charge state $q=80$.}    
\label{fig:sigMN}
\end{figure}

As the results in Fig.~\ref{fig:sigVsNEIES} show that the case of the initial charge state $q=80$ is an interesting case, we focus on this case in the following. Figure~\ref{fig:sigMN} presents the NERER cross section $\sigma_f$ for each recombined electron orbital of $M$ and $N$ shells for the initial charge state $q=80$. Here, three values of the energy of the incident electron are selected: $1$ eV, $100$ eV, and $10$ keV. For each recombined electron orbital, the NERER cross section $\sigma_f$ decreases with increasing the incident electron energy. For a given energy of the incident electron, and each subshell of a given orbital angular momentum quantum number $l_f$ and total angular momentum quantum number $j_f$ of the recombined electron orbital, the NERER cross section $\sigma_f$ decreases with increasing the principal quantum number $n_f$. This is easy to be understood, as a smaller $n_f$ means that the electron cloud is closer to the nucleus. This behavior shows the effect of the coupling to the atomic shells on the nuclear excitation.

It is shown in Fig.~\ref{fig:sigMN} that, for the $M1$ nuclear transition, the recombined channels with the highest efficiencies are the $s_{1/2}$ orbitals, and in each $n_f$, NERER cross section $\sigma_f$ decreases with increasing $l_f$ and $j_f$. For the $E2$ nuclear transition, the recombined channels with the highest efficiencies are the $p_{3/2}$ orbitals, and the $d_{3/2}$ ($f_{5/2}$) orbital shares similar result with the $d_{5/2}$ ($f_{7/2}$) orbital. This behavior clearly shows the effect of the selection rules for the process of NERER. Furthermore, we can observer from Figs.~\ref{fig:sigVsNEIES} and \ref{fig:sigMN} that, for $q=80$, recombining into $3s_{1/2}$ orbital is the main contribution for the summation cross section $\sigma_s$ for the $M1$ nuclear transition, while recombining into $3p_{3/2}$ orbital is the main contribution for the summation cross section $\sigma_s$ for the $E2$ nuclear transition.

\begin{figure}[htb!]    
\centering    
\includegraphics[width=1\linewidth]{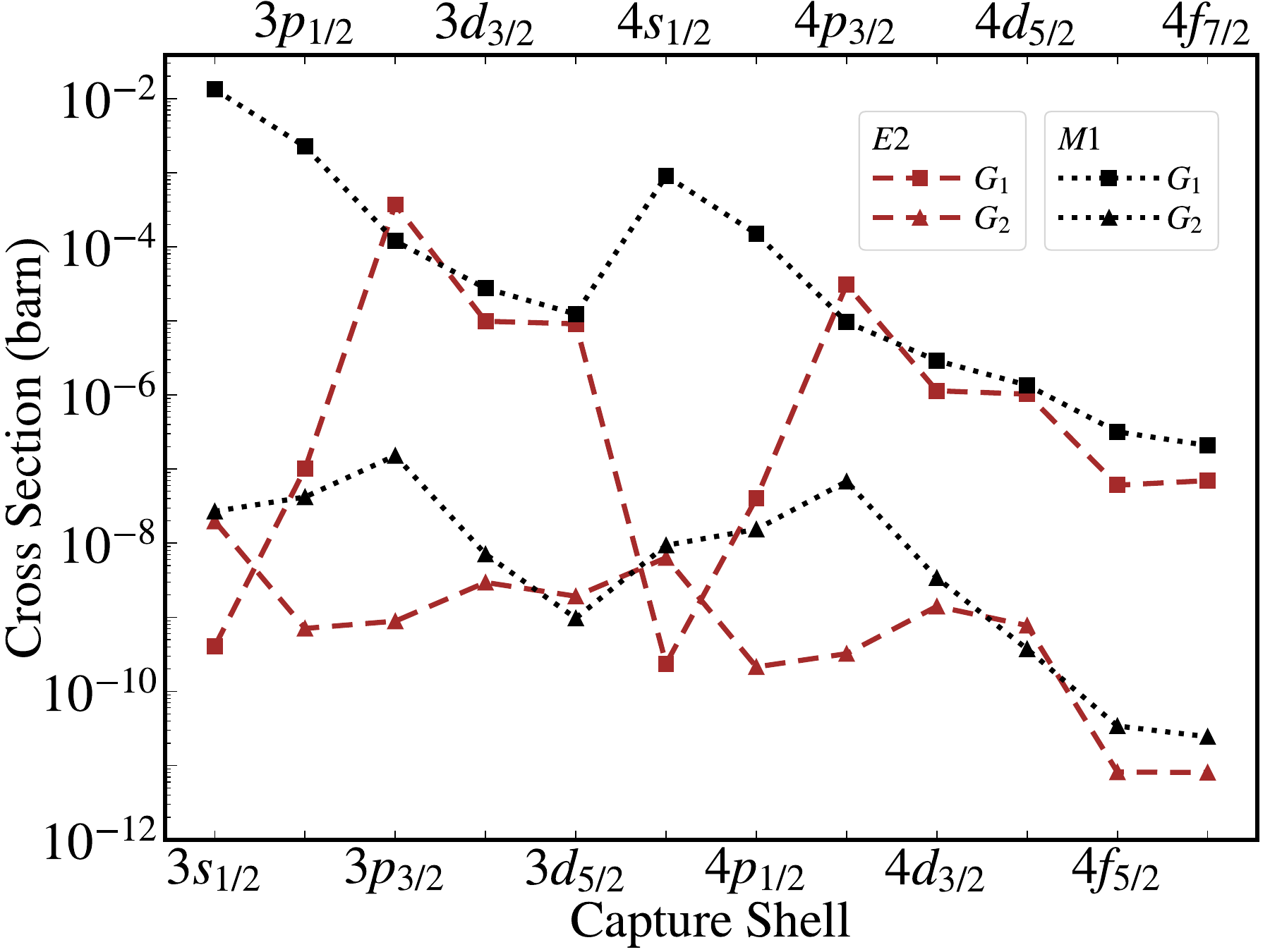}
\caption{The contribution from the $G_1$ term and the $G_2$ trem to the NERER cross section $\sigma_f$. Here the initial charge state $q=80$ and the incident electron energy of $100$ eV are considered.}  
\label{fig:sigMN100}
\end{figure}

It is also shown in Fig.~\ref{fig:sigMN} that, the dependence on the energy of the incident electron of the NERER cross section $\sigma_f$ for each recombined electron orbital is similar for orbitals of $M$ and $N$ shells, except $s_{1/2}$ orbitals for the $E2$ transition. In order to understand this behavior, we analyse the contribution from the $G_1$ term and the $G_2$ term which represent the two paths of the process. The results of the contribution from the $G_1$ term and the $G_2$ term for the incident electron energy of $100$ eV for $q=80$ are presented in Fig.~\ref{fig:sigMN100}. We can observe from Fig.~\ref{fig:sigMN100} that the main contribution for the NERER cross section is from the $G_1$ term, except $s_{1/2}$ orbitals for the $E2$ transition. We note that, for each recombined electron orbital and nuclear transition multipolarity, the absolute value of the contribution of the $G_{12}$ term is between the values of the contribution of the $G_1$ term and the $G_2$ term. Moreover, for the initial charge state $q=80$, the contribution from the $G_1$ term and the $G_2$ term for other incident electron energies share a similar behavior as the one of $100$ eV. This clearly shows that the path (i) is the main contribution for the NERER cross section in Fig.~\ref{fig:sigVsNEIES}. And the differences between the energy dependence of the path (i) and the path (ii) explain the above mentioned behavior of the energy dependence of the NERER cross section for $s_{1/2}$ orbitals for the $E2$ transition in Fig.~\ref{fig:sigMN}.

We have to mention here that, the direct excitation by NEEC for the production of $^{229m}$Th is also possible for highly-charged $^{229}$Th ions. Calculations on such scenario show that the resonance strength of NEEC can be on the level of $0.1$ b eV \cite{xu2025arXiv} at the resonant condition of sub-eV energy of the incident electron. We note that it is hard to make direct comparisons between NEEC and NERER, as NEEC is a resonant process while NERER is not. This means that only the electrons which satisfy the resonant condition on the level of the width of the state can contribute to NEEC. If we apply an electron beam with the energy spread on the level of eV or higher which could be easily accessed in experiments, the efficiency of NERER for the production of $^{229m}$Th could be more than about $2$ orders of magnitude higher than the one of NEEC. Moreover, we note that, when the incident electron energy falls into the resonant energy of a NEEC channel, enhancement for the $G_2$ contribution for the NERER cross section maybe expected. This means that for the part of the incident electron energy lower than the nuclear transition energy, the NERER cross section as a function of the incident electron energy should have some very narrow peaks. However, such peaks are not presented in Fig.~\ref{fig:sigVsNEIES}. This is because significant effect is expected only when the incident electron energy matches the resonant energy of a NEEC channel within a very narrow width, and only a few incident electron energies are studied in Fig.~\ref{fig:sigVsNEIES} as such effect is not the main focus in the present work.


In conclusion, we have put forward a nuclear excitation mechanism: nuclear excitation by radiative electron-ion recombination which is a high-order process that proceeds via a virtual electronic state. The photon emission compensates the energy mismatch between the free-bound electronic transition and the nuclear transition energies, thus there is no resonant condition imposed to the incident electron. The theoretical approach for NERER has been developed. Our calculations for the case of $8.4$ eV isomeric excitation of $^{229}$Th have shown that, with the coupling to the inner atomic shells for highly-charged ions, the NERER cross section can exceed the one of the known lower-order process of NEIES by more than one order of magnitude. Our findings offer a new pathway for nuclear excitation and the efficient production of the nuclear clock isomer $^{229m}$Th. Our analysis can be extended to other scenarios such as isomer depletion for energy storage solutions and isomer production for medical applications. Moreover, our findings also highlight the importance of high-order effects in nuclear excitation as well as the interplay between the atomic and nuclear degrees of freedom. It is an open question whether there exist more efficient mechanisms for nuclear excitation and important high-order effects which have not yet been realized so far.

\begin{acknowledgments}
  This work is supported by the National Natural Science Foundation of China (Grant No. 12475122), and by the Fundamental Research Funds for the Central Universities (Grant No. 010-63263118). 
\end{acknowledgments}

\bibliographystyle{apsrev-no-url-issn}
\bibliography{refsNERER26}{}

\clearpage

\noindent{\it{End matter -- Details of the theoretical approach.}} 

The Hamiltonian of the system can be written as
\begin{equation}
  H=H_{e}+H_{n}+H_{r}+H_{en}+H_{er}+H_{nr},
\end{equation} 
where $H_{e}$ is the Hamiltonian of the electron, $H_{n}$ is the Hamiltonian of the nucleus, and $H_{r}$ is the Hamiltonian of the radiation field. The Hamiltonian of the Coulomb interaction between the nucleus and the electron is given by
\begin{align}
H_{en}=\int\mathrm{d}\vec{r}_{n}\frac{\rho_{n}\left(\vec{r}_{n}\right)}{\left|\vec{r}_{e}-\vec{r}_{n}\right|},
\end{align}
where $\rho_{n}\left(\vec{r}_{n}\right)$ is the nuclear charge density. The Hamiltonian of the interaction between the electron and the radiation field is given by
\begin{align}
    H_{er}=-\vec{\alpha}\cdot\vec{A}\left(\vec{r}\right).
\end{align}
Here, $\vec{j}_{n}$ is the nuclear current density operator, $\vec{\alpha}$ is the vector of the Dirac matrices, and $\vec{A}$$\left(\vec{r}\right)$ is the vector potential of the radiation field. The Hamiltonian of the interaction between the nucleus and the radiation field is
\begin{align}
    H_{nr}=-\frac{1}{c}\int\vec{j}_{n}\left(\vec{r}\right)\cdot\vec{A}\left(\vec{r}\right)\mathrm{d}\vec{r}.
\end{align}

Following similar approaches for NEEC and NEET in Refs.~\cite{Palffy2006PRA,Palffy2007PRA,Arigapudi2012PRA}, we can develop the Feshbach projection operator for the
process, and
obtain the perturbation series for the transition operator from the Lippmann-Schwinger equation. This leads to the cross section in Eq.~\eqref{eq:sigmaF}, where $H_{magn}$ is given by
\begin{align}
    H_{magn}=-\frac{1}{c}\vec{\alpha}\cdot\int\mathrm{d}^{3}r_{n}\frac{\vec{j}_{n}\left(\vec{r}_{n}\right)}{\left|\vec{r}-\vec{r}_{n}\right|}.
\end{align}

In order to calculate the matrix elements, we expand the free electron wave function $\left|\vec{p}_{i}m_{s}\right>$ into partial wave series \cite{Zhang2022PRC,Liu2022PRC,Zhang2023FP,Rose1961AJP}
\begin{align}
    \left|\vec{p}_{i}m_{s}\right>
    &=\left|\vec{k}_{p_{i}}m_{s}\right>
    =\frac{4\pi}{k_{p_{i}}}\sqrt{\frac{\varepsilon_{i}+m_{e}c^{2}}{2\varepsilon_{i}}}
    \nonumber\\
    &\times
    \sum_{\kappa m}i^{l}\Omega_{\kappa m}^{\dagger}\left(\hat{k}_{p_{i}}\right)\chi_{m_{s}}e^{id_{\varepsilon_{i}\kappa}}
    \left|\varepsilon_{i}\kappa m\right>,
\end{align}
where $\vec{k}_{p_{i}}$ and $\varepsilon_{i}$ are respectively the wave vector and the total energy of the free electron, and $d_{\varepsilon_{i}\kappa}$ is the phase shift.
$\Omega_{\kappa m}$ is the spherical spinor and $\left|\varepsilon_{i}\kappa m\right>$ is the relativistic continuum electron wave function.
The details of $\Omega_{\kappa m}$, $\chi_{m_{s}}$ and $\left|\varepsilon_{i}\kappa m\right>$ can be found in Refs.~\cite{Zhang2022PRC,Liu2022PRC,Zhang2023FP}.

The treatment of the second term in $F_{2}$ can be adopted from Refs.~\cite{Dzyublik2020PRC,Borisyuk2019PRC}. With the partial wave expansion, it is straightforward to obtain the cross section Eq.~\eqref{eq:sigG}, with
\begin{align}
		G_{1}&
		= \sum_{\lambda_{f}L_{f}}\sum_{L}\sum_{\kappa}
		\left(\frac{4\pi}{p_{i}}\right)^{2}\frac{\varepsilon_{i}+m_{e}c^{2}}{2\varepsilon_{i}}\frac{2I_{f}+1}{2L+1}
		\left(2j_{f}+1\right)
		\nonumber\\
		&\times\sum_{n_{1}\kappa_{1}}\sum_{n_{1}^{'}\kappa_{1}^{'}\left(j_{1}^{'}=j_{1}\right)}
		\sum_{\lambda}
        \frac{\mathcal{T}_{\lambda L}^{\kappa_{f},\kappa_{1}}A_{\lambda_{f}L_{f}}^{\kappa_{1},\kappa}}{E_{f}^{n}+E_{f}^{e}-E_{1}^{e}-E_{i}^{n}+i\Gamma_{1}/2}
    \nonumber\\
    &\times
	\frac{\mathcal{T}_{\lambda L}^{\kappa_{f}\kappa_{1}^{'}*}A_{\lambda_{f}L_{f}}^{\kappa_{1}^{'},\kappa*}}{E_{f}^{n}+E_{f}^{e}-E_{1}^{e'}-E_{i}^{n}-i\Gamma_{1}^{'}/2}
    B\uparrow\left(\lambda L\right)
    ,
\end{align}

\begin{align}
    G_{2} & = \sum_{\lambda_{f}L_{f}}\sum_{\kappa}\sum_{L}
    \left(\frac{4\pi}{p_{i}}\right)^{2} \frac{\varepsilon_{i}+m_{e}c^{2}}{2\varepsilon_{i}}
    \frac{2I_{f}+1}{2L+1}\left(2j_{f}+1\right)
    \nonumber\\ &\times\sum_{n_{2}\kappa_{2}}\sum_{n_{2}^{'}\kappa_{2}^{'}\left(j_{2}^{'}=j_{2}\right)}\sum_{\lambda}\frac{A_{\lambda_{f}L_{f}}^{\kappa_{f},\kappa_{2}}\mathcal{T}_{\lambda L}^{\kappa_{2},\kappa}}
    {E_{i}^{n}+E_{i}^{e}-E_{f}^{n}-E_{2}^{e}+i\Gamma_{2}/2}
    \nonumber\\	&\times\frac{A_{\lambda_{f}L_{f}}^{\kappa_{f},\kappa_{2}^{'}*}
    \mathcal{T}_{\lambda L}^{\kappa_{2}^{'},\kappa*}}{E_{i}^{n}+E_{i}^{e}-E_{f}^{n}-E_{2}^{e'}-i\Gamma_{2}^{'}/2}
    B\uparrow\left(\lambda L\right)
    \nonumber\\
    &+
    \sum_{\lambda_{f}L_{f}}
        \sum_{\kappa}
		\sum_{L}
		\left(\frac{4\pi}{p_{i}}\right)^{2}\frac{\varepsilon_{i}+m_{e}c^{2}}{2\varepsilon_{i}}
		\frac{2I_{f}+1}{2L+1}
        \left(2j_{f}+1\right)
        \nonumber\\
        &\times
        \pi^{2}\rho^{2}(\varepsilon_{0})\left(\frac{4\pi}{p_{0}}\right)^{4}
		\left(\frac{\varepsilon_{0}+m_{e}c^{2}}{2\varepsilon_{0}}\right)^{2}
		\sum_{\kappa_{0}}
        \sum_{\kappa_{0}^{'}\left(j_{0}^{'}=j_{0}\right)}
        \sum_{\lambda}
		\nonumber\\
        & \times A_{\lambda_{f}L_{f}}^{\kappa_{f},\kappa_{0}}
		\mathcal{T}_{\lambda L}^{\kappa_{0},\kappa}
		A_{\lambda_{f}L_{f}}^{\kappa_{f},\kappa_{0}^{'}*}
		\mathcal{T}_{\lambda L}^{\kappa_{0}^{'},\kappa*}
        B\uparrow\left(\lambda L\right)
        \nonumber\\
    &+
    2\mathrm{Re}\left[\sum_{\lambda_{f}L_{f}}
		\sum_{n_{2}\kappa_{2}}
		\sum_{\kappa}
		\sum_{\lambda L}
		\sum_{\kappa_{0}\left(j_{0}=j_{2}\right)}
        \frac{\varepsilon_{i}+m_{e}c^{2}}{2\varepsilon_{i}}
        \left(\frac{4\pi}{p_{i}}\right)^{2}
        \right.
		\nonumber\\ 
        &\times 
        i\pi\rho\left(\varepsilon_{0}\right)
		\left(\frac{4\pi}{p_{0}}\right)^{2}\frac{\varepsilon_{0}+m_{e}c^{2}}{2\varepsilon_{0}}
        \frac{2I_{f}+1}{2L+1}
        \left(2j_{f}+1\right)
		\nonumber\\
		& 
        \left. \times \frac{A_{\lambda_{f}L_{f}}^{\kappa_{f},\kappa_{2}}
		\mathcal{T}_{\lambda L}^{\kappa_{2},\kappa} A_{\lambda_{f}L_{f}}^{\kappa_{f},\kappa_{0}*}
		\mathcal{T}_{\lambda L}^{\kappa_{0},\kappa*}
        }{E_{i}^{n}+E_{i}^{e}-E_{f}^{n}-E_{2}^{e}+i\Gamma_{2}/2}
        B\uparrow\left(\lambda L\right)
        \right]
        ,
        \\
        G_{12} & = \sum_{\lambda_{f}L_{f}}\sum_{n_{1}\kappa_{1}}\sum_{n_{2}\kappa_{2}}\sum_{\lambda L}\sum_{\kappa}
		\left(\frac{4\pi}{p_{i}}\right)^{2}\frac{\varepsilon_{i}+m_{e}c^{2}}{2\varepsilon_{i}}\frac{2I_{f}+1}{2L+1}
        \nonumber\\
        &\times
        \left(-1\right)^{j_{1}+2j-j_{2}-L+L_{f}}
		\left[\left(2j_{1}+1\right)\left(2j_{2}+1\right)\right]^{\frac{1}{2}}
		\left(2j_{f}+1\right)
        \nonumber\\
        &\times
		\left\{
		\begin{array}{cc c}
			j_{2}&L_{f}&j_{f}\\
			j_{1}&L&j
		\end{array}
		\right\}
        \times\frac{
        \mathcal{T}_{\lambda L}^{\kappa_{f},\kappa_{1}}
        A_{\lambda_{f}L_{f}}^{\kappa_{1},\kappa}
			}{E_{f}^{n}+E_{f}^{e}-E_{1}^{e}-E_{i}^{n}+i\Gamma_{1}/2}
		\nonumber\\
		&\times \frac{A_{\lambda_{f}L_{f}}^{\kappa_{f},\kappa_{2}*}
		\mathcal{T}_{\lambda L}^{\kappa_{2},\kappa*}}{E_{i}^{n}+E_{i}^{e}-E_{f}^{n}-E_{2}^{e}-i\Gamma_{2}/2}
        B\uparrow\left(\lambda L\right)
        \nonumber\\
    &+
    \sum_{\lambda_{f}L_{f}}
		\sum_{n_{1}\kappa_{1}}
		\sum_{\kappa_{0}}
		\sum_{\kappa}
		\sum_{\lambda L}
		\left(\frac{4\pi}{p_{i}}\right)^{2}\frac{\varepsilon_{i}+m_{e}c^{2}}{2\varepsilon_{i}}
        \frac{2I_{f}+1}{2L+1}
		\nonumber\\
		& \times i\pi\rho\left(\varepsilon_{0}\right)
		\left(\frac{4\pi}{p_{0}}\right)^{2}\frac{\varepsilon_{0}+m_{e}c^{2}}{2\varepsilon_{0}}
        \left\{
		\begin{array}{ccc}
			j_{0}&L_{f}&j_{f}\\
			j_{1}&L&j
		\end{array}
		\right\}
		\nonumber\\ 
        & 
        \times \left(-1\right)^{j_{1}+2j-j_{0}-L+L_{f}}
        \left[\left(2j_{1}+1\right)\left(2j_{0}+1\right)\right]^{\frac{1}{2}}
        \left(2j_{f}+1\right)
		\nonumber\\
		&
        \times 
         \frac{\mathcal{T}_{\lambda L}^{\kappa_{f},\kappa_{1}}
		A_{\lambda_{f}L_{f}}^{\kappa_{1},\kappa}
        A_{\lambda_{f}L_{f}}^{\kappa_{f},\kappa_{0}*}
		\mathcal{T}_{\lambda L}^{\kappa_{0},\kappa*}
        }{E_{f}^{n}+E_{f}^{e}-E_{1}^{e}-E_{i}^{n}+i\Gamma_{1}/2}
        B\uparrow\left(\lambda L\right)
        .
\end{align}
Here, $j$ is the total angular momentum quantum number, and $\lambda L$ is the transition multipolarity with $\lambda=M$ or $\lambda=E$. $\varepsilon_{0}=E_{p_{0}}^{e}+m_{e}c^{2}$, $E_{p_{0}}^{e}=E_{i}^{e}+E_{i}^{n}-E_{f}^{n}$, and $\rho\left(\varepsilon_{0}\right)$ is the density of states for electrons at energy $\varepsilon_{0}$.
$A_{\lambda_{f}L_{f}}^{\kappa_{f},\kappa_{2}}$, $A_{\lambda_{f}L_{f}}^{\kappa_{1},\kappa}$, $A_{\lambda_{f}L_{f}}^{\kappa_{f},\kappa_{0}}$,$\mathcal{T}_{\lambda L}^{\kappa_{f},\kappa_{1}}$, $\mathcal{T}_{\lambda L}^{\kappa_{2},\kappa}$, and $\mathcal{T}_{\lambda L}^{\kappa_{0}\kappa}$ are given by
\begin{align}
&\left<n_{f}\kappa_{f}m_{f},\lambda_{f}k_{f}L_{f}M_{f}\right|H_{er}\left|n_{2}\kappa_{2}m_{2}\right>
\nonumber\\
&= C\left(j_{2}L_{f}j_{f};m_{2}M_{f}m_{f}\right)
A_{\lambda_{f}L_{f}}^{\kappa_{f},\kappa_{2}},\\
&\left<n_{1}\kappa_{1}m_{1},\lambda_{f}k_{f}L_{f}M_{f}\right|H_{er}\left|\varepsilon_{i}\kappa m\right>\nonumber\\
&= 
C\left(jL_{f}j_{1};mM_{f}m_{1}\right)A_{\lambda_{f}L_{f}}^{\kappa_{1},\kappa},\\
&\left<n_{f}\kappa_{f}m_{f};\lambda_{f}k_{f}L_{f}M_{f}\right|H_{er}\left|\varepsilon_{0}\kappa_{0} m_{0}\right>\nonumber\\
&=C\left(j_{0}L_{f}j_{f};m_{0}M_{f}m_{f}\right)A_{\lambda_{f}L_{f}}^{\kappa_{f},\kappa_{0}},\\
&\left<N_{f}I_{f}M_{I_{f}},n_{f}\kappa_{f}m_{f}\right|H_{en/magn}\left|N_{i}I_{i}M_{I_{i}},n_{1}\kappa_{1}m_{1}\right>\nonumber\\
&= \sum_{L\mu}
\frac{\left<N_{f}I_{f}\right|\left|Q_{L}/M_{L}\right|\left|N_{i}I_{i}\right>
}{\sqrt{2I_{i}+1}}
C\left(I_{i}LI_{f};M_{I_{i}}\mu M_{I_{f}}\right)
\nonumber\\
&\times
\left(-1\right)^{\mu}
C\left(j_{1}L j_{f};m_{1},-\mu,m_{f}\right) \mathcal{T}_{E/M L}^{\kappa_{f}\kappa_{1}},
\\
&\left<N_{f}I_{f}M_{I_{f}},n_{2}\kappa_{2}m_{2}\right|H_{en/magn}\left|N_{i}I_{i}M_{I_{i}},\varepsilon_{i}\kappa m\right>\nonumber\\
&= \sum_{L\mu}
\frac{\left<N_{f}I_{f}\right|\left|Q_{L}/M_{L}\right|\left|N_{i}I_{i}\right>
}{\sqrt{2I_{i}+1}}
C\left(I_{i}LI_{f};M_{I_{i}}\mu M_{I_{f}}\right)\nonumber\\
&\times 
\left(-1\right)^{\mu}
C\left(jLj_{2};m,-\mu,m_{2}\right) \mathcal{T}_{E/M L}^{\kappa_{2},\kappa},
\end{align}
and
\begin{align}
&\left<N_{f}I_{f}M_{I_{f}},\varepsilon_{0}\kappa_{0}m_{0}\right|H_{en/magn}\left|N_{i}I_{i}M_{I_{i}},\varepsilon_{i}\kappa m\right>\nonumber\\
&= \sum_{L\mu}
\frac{\left<N_{f}I_{f}\right|\left|Q_{L}/M_{L}\right|\left|N_{i}I_{i}\right>
}{\sqrt{2I_{i}+1}}
C\left(I_{i}LI_{f};M_{I_{i}}\mu M_{I_{f}}\right)\nonumber\\
&\times 
\left(-1\right)^{\mu}
C\left(jLj_{0};m,-\mu,m_{0}\right) \mathcal{T}_{E/M L}^{\kappa_{0},\kappa}.
\end{align}

$\left<N_{f}I_{f}\right|\!\left|Q_{L}/M_{L}\right|\!\left|N_{i}I_{i}\right>$ is the nuclear reduced matrix elements, which can be connected to the nuclear reduced  transition probabilities by $B\uparrow(E/ML)=\left|\left<N_{f}I_{f}\right|\!\left|Q_{L}/M_{L}\right|\!\left|N_{i}I_{i}\right>\right|^{2}/(2I_{i}+1)$. $A_{\lambda_{f}L_{f}}^{\kappa_{f},\kappa_{2}}$, $A_{\lambda_{f}L_{f}}^{\kappa_{1},\kappa}$, $A_{\lambda_{f}L_{f}}^{\kappa_{f},\kappa_{0}}$, $\mathcal{T}_{\lambda L}^{\kappa_{f},\kappa_{1}}$, $\mathcal{T}_{\lambda L}^{\kappa_{2},\kappa}$ and $\mathcal{T}_{\lambda L}^{\kappa_{0}\kappa}$ take the follow forms
\begin{align}
    A_{ML_{f}}^{\kappa_{a},\kappa_{b}}
    &=
    i\sqrt{\frac{4\pi ck_{f}}{R}}
    \left(-1\right)^{L_{f}+\frac{1}{2}+j_{b}}\left(\kappa_{b}+\kappa_{a}\right)
    \nonumber\\
    & \times \left[\frac{\left(2j_{b}+1\right)\left(2l_{a}^{-}+1\right)\left(2l_{b}+1\right)\left(2L_{f}+1\right)}{4\pi L_{f}\left(L_{f}+1\right)}
    \right]^{\frac{1}{2}}
    \nonumber\\
    &\times
    \left(
    \begin{array}{ccc}
        L_{f}&l_{a}^{-}&l_{b}\\
			0&0&0
    \end{array}
    \right)
    \left\{
    \begin{array}{ccc}
	j_{a}&j_{b}&L_{f}\\
		l_{b}&l_{a}^{-}&\frac{1}{2}
    \end{array}
    \right\} R_{ML_{f}} ^{\kappa_{a},\kappa_{b}},
\end{align}
\begin{align}
    A_{EL_{f}}^{\kappa_{a},\kappa_{b}}&=
    i\sqrt{\frac{4\pi ck_{f}}{R}}		(-1)^{l_{a}^{-}+l_{b}+j_{b}+\frac{1}{2}}
    \left(
    \begin{array}{ccc}
        l_{a}&L_{f}&l_{b}\\
        0&0&0
    \end{array}
    \right)
    \nonumber\\
    &\times
    \left\{
    \begin{array}{ccc}
	l_{a}&L_{f}&l_{b}\\
		j_{b}&\frac{1}{2}&j_{a}
    \end{array}	\right\} \left[\frac{\left(2j_{b}+1\right)\left(2l_{b}+1\right)\left(2l_{a}+1\right)}{4\pi}\right]^{\frac{1}{2}}
    \nonumber\\
    &\times
    R_{EL_{f}}^{\kappa_{a},\kappa_{b}},
\end{align}
\begin{align}
    \displaybreak[2] \mathcal{T}_{ML}^{\kappa_{a},\kappa_{b}}
    &=
	\frac{4\pi}{2L+1}\sqrt{\frac{L+1}{L}}
		\sqrt{\frac{2I_{i}+1}{2I_{f}+1}}
        \left(-1\right)^{L+\frac{1}{2}+j_{b}}
        \nonumber\\
		&
		\times
        \left[
        \frac{\left(2j_{b}+1\right)\left(2l_{a}^{-}+1\right)\left(2l_{b}+1\right)\left(2L+1\right)}{4\pi L\left(L+1\right)}
		\right]^{\frac{1}{2}}
        \nonumber\\
        &
        \times
        \left(\kappa_{a}+\kappa_{b}\right)
		\left(
		\begin{array}{ccc}
			L&l_{a}^{-}&l_{b}\\
			0&0&0
		\end{array}
		\right)
		\left\{
		\begin{array}{ccc}
			j_{a}&j_{b}&L\\
			l_{b}&l_{a}^{-}&\frac{1}{2}
		\end{array}
		\right\}
		M_{ML}^{\kappa_{a},\kappa_{b}},
\end{align}
and
\begin{align}
    \mathcal{T}_{EL}^{\kappa_{a},\kappa_{b}}
		&=
		\frac{4\pi}{2L+1}
        \sqrt{\frac{2I_{i}+1}{2I_{f}+1}}
		(-1)^{-j_{b}-L-\frac{1}{2}}
        \nonumber\\
        &\times
		\left[\frac{(2l_{a}+1)(2L+1)(2l_{b}+1)(2j_{b}+1)}{4\pi}\right]^{\frac{1}{2}}
        \nonumber\\
        &
        \times
        \left(
		\begin{array}{ccc}
			l_{a}&L&l_{b}\\
			0&0&0\\
		\end{array}
		\right)
		\left\{
		\begin{array}{ccc}
			j_{a}&L&j_{b}\\
			l_{b}&\frac{1}{2}&l_{a}
		\end{array}
		\right\}
        M_{EL}^{\kappa_{a},\kappa_{b}},
\end{align}
where $l_a$ is the orbital angular momentum quantum number and $l_{a}^{-}\!=\!2j_{a}-l_{a}$. $C$ is the Clebsch-Gordan coefficient, $\binom{...}{...}$ is the Wigner $3j$-symbol and $\genfrac{\{}{\}}{0pt}{}{...}{...}$ is the the Wigner $6j$-symbol \cite{Edmonds1996PUP}. The radial integrals $R_{ML_{f}}^{\kappa_{a},\kappa_{b}}$, $R_{EL_{f}}^{\kappa_{a},\kappa_{b}}$, $M_{ML_{f}}^{\kappa_{a},\kappa_{b}}$ and $M_{EL_{f}} ^{\kappa_{a},\kappa_{b}}$ are given by
\begin{align}
    R_{ML_{f}}^{\kappa_{a},\kappa_{b}}=\int\mathrm{d}rr^{2}j_{L_{f}}(k_{f}r)\left[g_{n_{a}\kappa_{a}}(r)\right.&f_{n_{b}\kappa_{b}}(r)\nonumber\\
    +f_{n_{a}\kappa_{a}}&(r)\left.g_{n_{b}\kappa_{b}}(r)\right],
\end{align}
\begin{align}
    R_{EL_{f}}^{\kappa_{a}, \kappa_{b}}&=
    \sqrt{\frac{L_{f}+1}{L_{f}(2L_{f}+1)}}\left(L_{f}I_{L_{f}-1}^{-}-\left(\kappa_{a}-\kappa_{b}\right)I_{L_{f}-1}^{+}\right)
    \nonumber\\
    &\left.+\sqrt{\frac{L_{f}}{(L_{f}+1)(2L_{f}+1)}}\left(\left(L_{f}+1\right)I_{L_{f}+1}^{-}
    \right.\right.
    \nonumber\\
    &\qquad\qquad\qquad\qquad\qquad
    \left.
    +(\kappa_{a}-\kappa_{b})I_{L_{f}+1}^{+}\right),
\end{align}
\begin{align}
    M_{ML_{f}}^{\kappa_{a},\kappa_{b}}&=\int\mathrm{d}rr^{-L+1}\left[g_{n_{a}\kappa_{a}}(r)f_{n_{b}\kappa_{b}}(r)
    \right.
    \nonumber\\
    &\left.\qquad\qquad\qquad\qquad\quad
    +f_{n_{a}\kappa_{a}}(r)g_{n_{b}\kappa_{b}}(r)\right],
\end{align}
and
\begin{align}
        M_{EL_{f}} ^{\kappa_{a},\kappa_{b}}
		&=\int\mathrm{d}rr^{-L+1}\left[f_{n_{a}\kappa_{a}}(r)f_{n_{b}\kappa_{b}}(r)\right.
        \nonumber\\
        &
        \left.\qquad\qquad\qquad\qquad\qquad
        +g_{n_{a}\kappa_{a}}(r)g_{n_{b}\kappa_{b}}(r)\right].
\end{align}
Here, $I_{L}^{-}$ and $I_{L}^{+}$ are given by
\begin{align}
		I_{L}^{-}&=\int\mathrm{d}rr^{2}j_{L}(k_{f}r)\left[g_{n_{a}\kappa_{a}}(r)f_{n_{b}\kappa_{b}}(r)\right.
        \nonumber\\
        &\left.\qquad\qquad\qquad\qquad\qquad\quad\;
        -f_{n_{a}\kappa_{a}}(r)g_{n_{b}\kappa_{b}}(r)\right],
\end{align}
and
\begin{align}
		I_{L}^{+}&=\int\mathrm{d}rr^{2}j_{L}(k_{f}r)\left[g_{n_{a}\kappa_{a}}(r)f_{n_{b}\kappa_{b}}(r)
        \right.
        \nonumber\\
        &
        \left.\qquad\qquad\qquad\qquad\qquad\quad\;
        +f_{n_{a}\kappa_{a}}(r)g_{n_{b}\kappa_{b}}(r)\right],
\end{align}
where $j_{L}$ is the spherical Bessel function. The functions $g$ and $f$ are the large and small radial components of the electron wave function \cite{Zhang2022PRC,Liu2022PRC,Zhang2023FP,Palffy2006PRA,Palffy2007PRA}.

\end{document}